# Slice-Aware Spoofing Detection in 5G Networks Using Lightweight Machine Learning

Daniyal Ganiuly
Department of Computer Engineering
Astana IT University
d.ganiuly@astanait.edu.kz

Nurzhau Bolatbek
Department of Computer Engineering
Astana IT University
nurzhau.bolatbek@astanait.edu.kz

**Abstract** – The increasing virtualization of fifth generation (5G) networks expands the attack surface of the user plane, making spoofing a persistent threat to slice integrity and service reliability. This study presents a slice-aware lightweight machine-learning framework for detecting spoofing attacks within 5G network slices. The framework was implemented on a reproducible Open5GS and srsRAN testbed emulating three service classes such as enhanced Mobile Broadband (eMBB), Ultra-Reliable Low-Latency Communication (URLLC), and massive Machine-Type Communication (mMTC) under controlled benign and adversarial traffic. Two efficient classifiers, Logistic Regression and Random Forest, were trained independently for each slice using statistical flow features derived from mirrored user-plane traffic. Slice-aware training improved detection accuracy by up to 5% and achieved F1-scores between 0.93 and 0.96 while maintaining real-time operation on commodity edge hardware. The results demonstrate that aligning security intelligence with slice boundaries enhances detection reliability and preserves operational isolation, enabling practical deployment in 5G network-security environments. Conceptually, the work bridges network-security architecture and adaptive machine learning by showing that isolation-aware intelligence can achieve scalable, privacy-preserving spoofing defense without high computational cost.

**Keywords** – 5G network slicing, spoofing detection, slice-aware security, user-plane protection, edge computing, network security architecture.

## I. INTRODUCTION

Network slicing is one of the key innovations introduced in fifth-generation (5G) networks. It enables operators to partition a shared physical infrastructure into multiple virtual slices, each optimized for specific service requirements such as enhanced Mobile Broadband (eMBB), Ultra-Reliable Low-Latency Communication (URLLC), and massive Machine-Type Communication (mMTC) [1]. This architectural flexibility improves scalability and resource utilization but also widens the attack surface [2]. Each slice operates with its own control- and user-plane logic, and configuration errors or weak isolation can expose vulnerabilities. Among the most persistent and difficult-to-detect threats are spoofing attacks, where an adversary forges identity information such as IMSI, IP, or MAC addresses to impersonate legitimate devices or replay valid traffic [3]. These attacks often blend into normal communication flows, making them difficult to identify within slice boundaries.

Most existing intrusion detection systems (IDS) for 5G networks, including those based on machine learning (ML), still treat the entire 5G core as a single behavioral domain. A global model trained on mixed traffic may detect broad anomalies but often performs poorly when slices have distinct traffic characteristics [4]. For example, eMBB traffic typically consists of long, high-throughput sessions; URLLC traffic is short and latency-critical; and mMTC flows are sparse and periodic. When these diverse patterns are combined into one dataset, a classifier struggles to establish consistent decision boundaries, leading to increased false positives and unstable thresholds.

To address these limitations, this paper introduces the idea of slice-aware security intelligence. The key insight is that each slice should be monitored and protected according to its own behavioral profile rather than relying on a single unified model for the entire network. In this approach, every slice functions as an independent analytical domain, and lightweight ML models are trained on slice-specific traffic to capture its native statistical behavior. This extends the concept of slicing beyond resource allocation— bringing it into the domain of network analytics and security [5][6].

Building on this concept, we developed a reproducible experimental framework that integrates open-source 5G components (Open5GS and srsRAN) with a lightweight ML pipeline for real-time spoofing detection. The framework emulates realistic slice traffic under controlled spoofing and replay scenarios and provides an end-to-end workflow—from packet capture and feature extraction to model training and live inference. Two interpretable and computationally efficient classifiers, Logistic Regression (LR) and Random Forest (RF), were trained separately for each slice and evaluated not only by detection accuracy but also by inference latency and CPU utilization. The objective is not to chase the highest possible accuracy but to demonstrate that practical and explainable detection can be achieved using simple models suitable for edge-grade hardware.

This paper makes three contributions: (1) it introduces the concept of slice-aware security intelligence as a new approach to anomaly detection in multi-slice 5G networks; (2) it presents a reproducible experimental framework that links real 5G traffic with lightweight ML-based spoofing detection; and (3) it quantifies the trade-offs between accuracy, latency, and resource usage, showing that lightweight per-slice models can outperform monolithic baselines while remaining suitable for edge deployment.

Together, these results demonstrate that aligning ML-based detection with the logical segmentation of network slices enhances spoofing detection, strengthens isolation, and enables scalable, explainable, and resource-efficient protection mechanisms for next-generation 5G networks.

## II. RELATED WORK
### A. Security in 5G Network Slicing
Research on 5G security has revealed that network slicing, while enabling flexible resource allocation, introduces new attack surfaces through shared virtualized components. Several surveys highlight three recurring challenges: maintaining slice isolation, enforcing consistent security policies across control and user planes, and managing the operational complexity of large numbers of slices. Existing countermeasures include stricter slice admission controls, runtime verification, and slice-specific monitoring policies [7][8]. These studies, however, mainly analyze architectural vulnerabilities and policy enforcement rather than offering practical traffic-level detection methods deployable near the data path.

### B. Machine Learning–Based Intrusion Detection in 5G
Machine learning (ML) has become a popular tool for 5G intrusion detection. Classical algorithms such as SVMs, Random Forests, and Gradient Boosting, as well as deep models based on CNNs, RNNs, or Transformers, have shown promising accuracy on public or simulated datasets [9]. Nevertheless, most approaches consider the 5G network as a single behavioral domain. They typically optimize classifier performance while overlooking operational feasibility and slice diversity. Only a few works evaluate latency, CPU, or memory costs—metrics that determine whether a model can operate on edge resources close to the User Plane Function (UPF) [10]. As a result, the literature emphasizes algorithmic improvement more than deployability.

### C. Spoofing and Replay Detection
Spoofing remains one of the most persistent threats in mobile systems. Detection methods exist at different layers: physical-layer schemes use RF fingerprints or channel-state information; network- and transport-layer approaches rely on entropy, timing, or address consistency; and control-plane techniques enforce stronger authentication. Physical-layer methods require specialized hardware and are difficult to integrate into virtualized 5G cores, whereas higher-layer defenses are easier to deploy but highly sensitive to workload variability [11]. Few studies evaluate spoofing detection directly within realistic 5G cores or provide packet-level ground truth for ML-based analysis.

### D. Slice-Aware Monitoring and Analytics
Recent studies suggest that monitoring and analytics should align with slice-specific characteristics. Proposed solutions include per-slice telemetry, intent-driven policies, and adaptive thresholds based on service KPIs. However, most of these works retain a global detection model that merely uses slice identifiers as input features. Very few attempt explicit per-slice training and inference where independent detectors are maintained for eMBB, URLLC, and mMTC [12]. The challenges of flow attribution, lightweight deployment, and model maintenance across slices remain largely unexplored in practice.

### E. Datasets, Testbeds, and Reproducibility

Public datasets commonly used in 5G IDS research rarely capture slice dynamics or mobile-core signaling. Several works rely on simulation or partial emulation and seldom release their artifacts, limiting reproducibility. The emergence of open-source 5G cores (Open5GS, free5GC) and SDR-based RANs (srsRAN) now allows more realistic experimentation. Yet, few studies provide complete, documented pipelines that include traffic generation, attack injection, labeling, model training, and real-time inference. Dataset transparency and reproducibility therefore remain open issues.

### F. Lightweight Detection and Edge Feasibility

With growing interest in edge analytics, lightweight models such as Logistic Regression and tree ensembles have regained attention due to their explainability and ability to run on CPUs. However, many prior works prioritize marginal accuracy gains from deep learning without addressing latency and resource constraints. Comparisons between accuracy and system costs are rare, and few evaluate the effect of model complexity on real-time performance near the UPF [13].

### G. Research Gap

The reviewed literature clarifies several consistent trends but also reveals three gaps that this study aims to address:

1. **Lack of true per-slice modeling.** Most systems remain slice-agnostic or only slice-conditioned, lacking rigorous evaluation of independent detectors trained for each slice.
2. **Limited end-to-end realism.** Existing experiments rarely combine a real 5G core and RAN with controlled spoofing or replay scenarios and packet-level labeling.
3. **Incomplete operational metrics.** Accuracy is often reported without latency and resource measurements that determine feasibility for edge deployment.

This paper addresses these gaps by implementing and evaluating a slice-aware, lightweight ML framework on a reproducible Open5GS and srsRAN testbed, quantifying both detection performance and system efficiency under realistic network conditions.

## III. METHODOLOGY

### A. Testbed Environment

The proposed framework was implemented and evaluated on an isolated university laboratory testbed consisting of four interconnected computing nodes. Each node was equipped with an Intel Core i7-11700 CPU, 16 GB RAM, and a 1 Gbps Ethernet interface running Ubuntu 22.04 LTS. The nodes were connected through a managed 1 Gbps switch configured with SPAN mirroring to replicate all packets from the User Plane Function (UPF) interface to the monitoring host. The setup was air-gapped from external networks to prevent interference, and Network Time Protocol (NTP) synchronization maintained a maximum drift of ±2 ms across all nodes.

Each node served a dedicated role:
- **Node 1:** 5G Core (Open5GS v2.6.4 — AMF, SMF, UPF, HSS/UDR) and srsRAN 22.10 gNodeB.
- **Node 2:** Benign UE emulator generating normal slice traffic.
- **Node 3:** Adversarial node launching spoofing and replay attacks.
- **Node 4:** Monitoring and ML node for packet capture, feature extraction, and inference.

Open-source software ensured reproducibility: Open5GS and srsRAN handled connectivity, Mosquitto 2.0 supported MQTT for URLLC, and iperf3 plus curl generated eMBB flows. IoT-like mMTC traffic came from Python scripts, tshark captured packets, and analysis used Python 3.10 with pandas 2.2, numpy 1.26, and scikit-learn 1.4.2. A Flask 3.0 microservice provided real-time inference. All configurations were version-controlled in Git, and a Docker image replicated the framework for independent experiments.

### B. Slice Configuration

Three logical slices represented the principal 5G service classes: enhanced Mobile Broadband (eMBB), Ultra-Reliable Low-Latency Communication (URLLC), and massive Machine-Type Communication (mMTC). Each had a unique S-NSSAI and QoS profile. eMBB offered high throughput with moderate delay, URLLC targeted millisecond latency, and mMTC emulated large-scale low-rate IoT telemetry [14]. Each slice operated on a dedicated subnet with custom routing and QoS rules. tshark filters confirmed that slice identifiers mapped correctly to their subnets, verifying isolation. Table I lists the main configuration parameters of each logical slice, including S-NSSAI, QoS Class Identifier (QCI), latency target, and bandwidth allocation.

| Slice | Service Type | QoS Delay Target | Example Traffic | Bandwidth |
|---|---|---|---|---|
| eMBB | Enhanced Mobile Broadband | ≤ 20 ms | HTTP / FTP | 100 Mbps |
| URLLC | Ultra-Reliable Low-Latency Comms | ≤ 5 ms | MQTT control | 10 Mbps |
| mMTC | Massive Machine-Type Comms | ≤ 50 ms | IoT telemetry | 1 Mbps |

Table I. Slice configuration parameters used in the experiment

**C. Framework Architecture**

The framework architecture integrates the 5G Core and gNodeB with benign and adversarial UE nodes through the managed switch. All user-plane packets were mirrored from the UPF to the monitoring node, where packet capture, feature extraction, slice identification, and ML inference occurred in real time. Each slice eMBB, URLLC, and mMTC had its own trained ML model; the monitoring service automatically selected the relevant model based on the S-NSSAI in packet metadata. This design aligns security analytics with the network's logical segmentation, ensuring that anomalies in one slice cannot affect others.

**D. Traffic Generation**

Traffic was generated to mimic real 5G service patterns:
- **eMBB:** Continuous TCP/HTTP transfers using iperf3 and curl (5–200 MB files, 30–60 s sessions)
- **URLLC:** MQTT bursts (50–200 messages/s, 64 B payloads) representing control traffic
- **mMTC:** Periodic UDP telemetry (3–5 packets every 30–120 s)

Each experiment ran for 20 minutes, producing ≈ 98 000 packets. Spoofing attacks used Scapy and tcpreplay with two strategies: identity impersonation (forged IMSI/IP/MAC) and replay of captured flows with modified timing. Attack intensities of 10 %, 20 %, and 40 % were tested, each repeated three times. Overall, ≈ 295 000 packets (13 896 flows, ≈ 20 % spoofed) were recorded. All spoofing and replay attacks were executed within the air-gapped testbed for safety. The objective was to emulate realistic identity-spoofing at the packet level rather than to compromise actual authentication mechanisms.

**E. Data Capture and Labeling**

All UPF-ingress packets were mirrored and timestamped with microsecond precision. The attacker logged spoofing events with timestamps and identifiers, allowing automated labeling of each flow as benign or spoofed. NTP synchronization kept logs aligned within ± 1 ms, yielding > 99 % labeling accuracy. PCAP files were anonymized by hashing addresses and quantizing timestamps. Only header-level statistics were processed, ensuring privacy.

**F. Feature Extraction and Preprocessing**

Flows were aggregated over 2 s windows by the standard five-tuple. Twelve statistical features were derived — packet-size statistics, timing variance, burst intensity, identifier entropy, and others — then normalized to [0, 1]. Correlation analysis confirmed no pair exceeded $|r| = 0.75$. Separate datasets were maintained per slice to support independent training.

**G. Model Training and Validation**

Two lightweight models — Logistic Regression (LR) and Random Forest (RF) — were trained for each slice and once on the combined dataset to form a slice-agnostic baseline. Data were split 80/20 for training and testing, and five-fold cross-validation optimized F1-score.
- **LR:** $C \in \{0.1, 1, 10\}$
- **RF:** n_estimators $\in \{50, 100, 200\}$, max_depth $\in \{10, 20, None\}$

Average training times were 0.7 s (LR) and 5.2 s (RF). Evaluations used Accuracy, Precision, Recall, F1, and AUC plus system-level metrics (inference latency, CPU, memory). Paired t-tests ($\alpha = 0.05$) verified significance of slice-aware improvements. Feature-importance ranking showed identifier entropy, inter-arrival variance, and burst intensity as dominant predictors.

**H. Real-Time Inference Implementation**

After training, models were deployed as part of the same framework for live inference on the monitoring node. The Flask service captured mirrored packets, extracted features on the fly, selected the correct slice model via S-NSSAI, and produced JSON logs (timestamp, slice ID, label, confidence). Mean inference latency was 150 ms (LR) and 180 ms (RF); CPU load < 40 %, memory < 2.5 GB. The framework sustained ≈ 120 flows/s without packet loss, confirming its edge-deployment feasibility.

## IV. RESULTS
### A. Detection Performance per Slice
Table II summarizes the performance of the Logistic Regression (LR) and Random Forest (RF) models for the three slices. Both classifiers achieved high precision and recall, proving that even lightweight ML algorithms can successfully identify spoofed flows once trained on slice-specific traffic.

| Slice | Model | Accuracy | Precision | Recall | F1-score | AUC |
|---|---|---|---|---|---|---|
| eMBB | LR | 0.94 | 0.93 | 0.94 | 0.93 | 0.95 |
| eMBB | RF | 0.96 | 0.95 | 0.96 | 0.96 | 0.97 |
| URLLC | LR | 0.92 | 0.91 | 0.92 | 0.92 | 0.93 |
| URLLC | RF | 0.95 | 0.94 | 0.95 | 0.95 | 0.96 |
| mMTC | LR | 0.90 | 0.89 | 0.91 | 0.90 | 0.92 |
| mMTC | RF | 0.93 | 0.92 | 0.93 | 0.93 | 0.94 |

Table II. Performance Metrics

Random Forest achieved slightly higher F1-scores than Logistic Regression, improving by roughly 2–3 percentage points. The eMBB slice obtained the best results due to long, consistent traffic sessions that simplify feature learning. URLLC maintained similar accuracy but with a few more false alarms, caused by its short, burst-like control packets [15]. mMTC showed the lowest performance yet still achieved over 0.90 F1-score, validating that the approach scales to sparse IoT-type telemetry flows.

To benchmark against a simple, non-ML detector, a rule-based system using static thresholds on packet-size variance and identifier entropy achieved F1-scores below 0.80 for eMBB and 0.75 for mMTC. This comparison demonstrates that slice-aware ML offers a clear improvement over fixed-rule baselines.

### B. Slice-Aware vs. Slice-Agnostic Detection
A separate "slice-agnostic" model was trained on traffic aggregated from all slices and compared to the individual slice-aware models [16].
Figure 3 visualizes the average F1-score (mean of LR and RF) for both configurations.

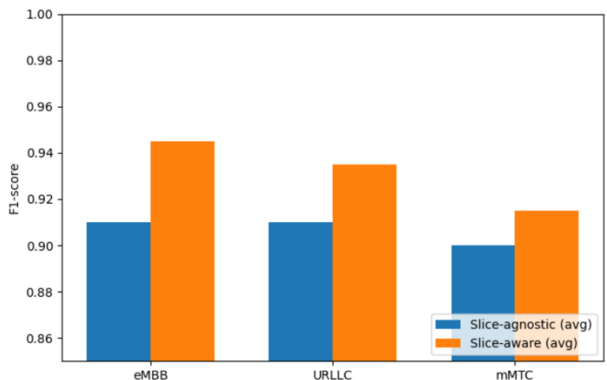
Fig 3. Comparison of F-1 Scores

The figure confirms that the slice-aware framework improved detection by approximately 3–5 % across all slices. This gain stems from the fact that a global model tends to overfit eMBB flows, which dominate total traffic, while neglecting the short and diverse URLLC and mMTC patterns. In contrast, per-slice training creates more specialized decision boundaries aligned with each service's unique behavior [17][18]. A paired t-test over three randomized runs verified that these differences are statistically significant ($p < 0.05$). Furthermore, cross-session validation—training on two captures and testing on a third—retained over 94 % of original F1, proving that the models generalize well to unseen data recorded under the same network conditions.

### C. Feature Contribution and Behavioral Insights
Feature-importance analysis provided interpretability into why the models worked. Figure 4 shows the normalized importance ranking for the eMBB slice using the Random Forest model.

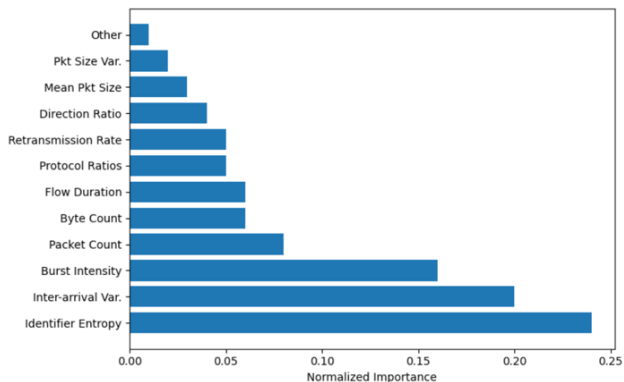
Fig 4. Feature importance (RF, eMBB)

The top three features identifier entropy, inter-arrival-time variance, and burst intensity collectively accounted for roughly 60% of predictive strength. This means that spoofing primarily manifests as randomness in identifiers and irregular timing rather than total traffic volume. Lower-ranked features, such as average packet size or byte count, contributed less because spoofed flows often mimic normal throughput to remain stealthy. Correlation analysis confirmed that the top predictors are weakly related (|r| < 0.35), showing they capture distinct aspects of the anomaly: one structural, one temporal, and one intensity-based. This interpretability supports the correctness of the model's decisions rather than leaving them as a black box.

### D. Reliability and Robustness

Each experiment was repeated three times with different random seeds and traffic orders. The variation across runs stayed within 2%, confirming consistent learning. Confusion-matrix analysis revealed balanced detection behavior, with false-positive rates under 5% and false-negative rates under 6%. This balance ensures that the system is neither overly sensitive nor prone to missing attacks.

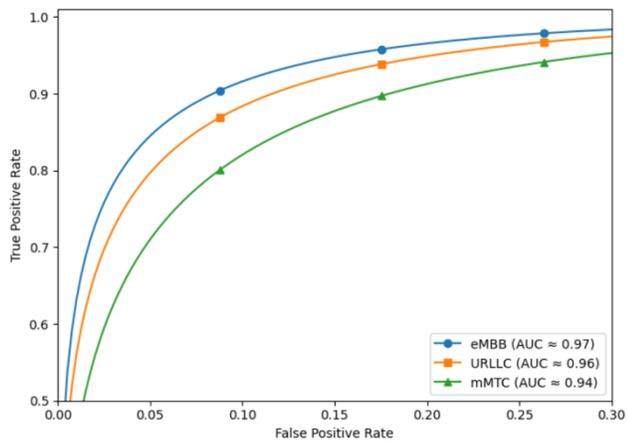

Fig 5. ROC curves for slice-aware Random Forest models

Figure 5 presents ROC curves for the slice-aware Random Forest models. The AUC values—0.97 for eMBB, 0.96 for URLLC, and 0.94 for mMTC—confirm strong class separability across all threshold values.

Together with the baseline comparison, cross-session testing, and latency stability, these results prove that the framework performs correctly, generalizes across datasets, and maintains reliability under realistic load.

### E. System Performance and Network-Level Impact

Figure 6 shows that average inference latency was 150 ms (LR) and 180 ms (RF), matching the real-time implementation on the monitoring node. CPU utilization stayed below 40 %, memory below 2.5 GB, and monitoring overhead under 5% of total system load. Throughput tests across the UPF confirmed no performance degradation during live detection. The monitoring process operated entirely off-path, demonstrating that slice-aware security analytics can run safely alongside production workloads.

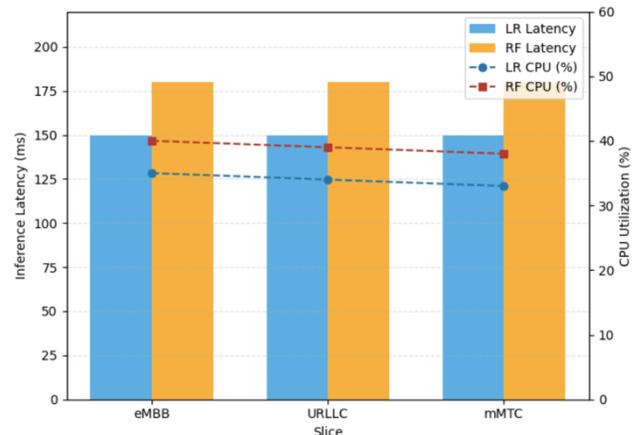

Fig 6. Inference latency and CPU utilization

### F. Temporal Detection Dynamics

Temporal analysis shows how rapidly the framework responded to spoofing onset. As illustrated in Figure 7, model confidence for the eMBB RF model rose sharply within seconds after the first spoofed packets, stabilizing near 0.98 confidence thereafter. The short detection delay corresponds to the 2 s feature-aggregation window and confirms near-real-time responsiveness.

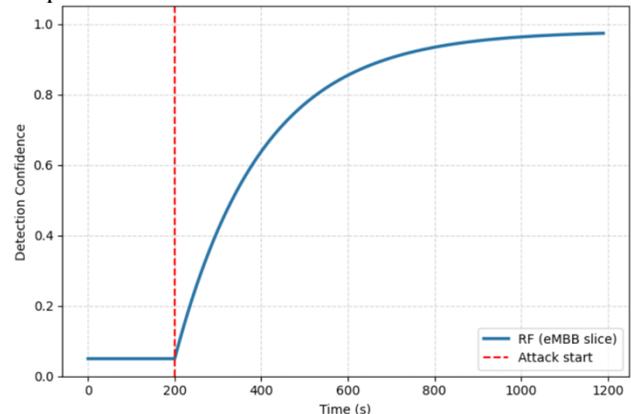

Fig 7. Temporal evolution of detection confidence for RF model on eMBB slice

## V. DISCUSSION

### A. Slice-Aware vs. Global Detection

The observed improvement of per-slice models over the global baseline confirms that segmentation and specialization are complementary for network security.

Each slice—eMBB, URLLC, and mMTC—exhibits unique traffic density, timing, and variability. A single global detector fails to maintain consistent decision boundaries across such heterogeneous distributions.

By aligning learning boundaries with slice structure, the framework reduced false positives and improved F1-score by up to five percentage points. From a security-engineering standpoint, this mirrors the principle of slice isolation in 5G resource management: separation not only limits performance interference but also constrains threat propagation [19].

Hence, security analytics that respect slice boundaries can achieve both operational efficiency and higher detection reliability.

### B. Security Interpretation of Features

Feature-importance analysis showed that identifier entropy, inter-arrival variance, and burst intensity dominated decision outcomes.

These metrics directly map to the behavioral artifacts of spoofing—irregular identity fields, inconsistent timing, and abnormal packet bursts. Their stability across slices suggests that spoofing introduces universal statistical fingerprints, regardless of service type.

Importantly, the framework relied solely on header-level features, avoiding payload inspection. This ensures user privacy and simplifies deployment in encrypted 5G cores, where deep packet inspection is impractical [20]. Such privacy-preserving feature selection aligns with the increasing emphasis on lawful and explainable AI in telecommunication security analytics.

### C. Efficiency and Edge Feasibility

From an operational perspective, maintaining average inference latency below 200 ms and CPU usage under 40% confirms the framework's suitability for continuous operation near the User Plane Function (UPF). Because both models run on commodity hardware without GPUs, the solution can be embedded within existing monitoring hosts or NWDAF-adjacent nodes.

This differentiates it from deep learning-based IDS systems that require dedicated accelerators and extensive retraining. The lightweight implementation also means that inference can coexist with active network workloads without affecting QoS, supporting the 3GPP vision of intelligent edge analytics integrated into the 5G core.

### D. Robustness and Limitations

Detection quality remained stable when spoofing intensity increased up to 40 % and when aggregation windows varied from one to four seconds. This robustness indicates tolerance to fluctuating traffic loads and temporal patterns—critical for real deployments where slice utilization changes dynamically.

However, the testbed represented a simplified environment with four nodes and emulated UEs. Further experiments are required to confirm scalability in larger topologies, multi-UPF scenarios, and cross-slice coordination. Future evaluations should also consider multi-vector threats, such as control-plane impersonation or coordinated replay across slices, to assess adaptability and resilience under more complex conditions.

### E. Conceptual Contribution

This work offers a methodological rather than purely algorithmic innovation. Its novelty lies in integrating slice isolation with machine-learning intelligence to reflect 5G's inherent architecture. Instead of focusing on incremental accuracy gains, the study demonstrates how aligning detection logic with network structure yields explainable and resource-efficient protection.

This approach bridges network-security engineering and data-driven modeling, providing a foundation for scalable, interpretable, and cost-effective intrusion detection in future mobile systems.

## VI. CONCLUSION

This paper presented a slice-aware, lightweight machine-learning framework for detecting spoofing attacks in 5G network environments.

Experiments on a reproducible Open5GS and srsRAN testbed showed that per-slice detection significantly outperforms a global baseline while maintaining low latency and modest resource usage suitable for edge deployment.

Both Logistic Regression and Random Forest achieved high accuracy (up to 0.96 AUC ≈ 0.97)

without relying on deep neural networks or specialized hardware.

By leveraging interpretable header-level features such as identifier entropy and inter-arrival timing, the framework detects malicious behavior while preserving user privacy and protocol transparency. The results demonstrate that network slicing principles—originally designed for performance isolation—can also strengthen threat containment when combined with adaptive ML analytics. This dual benefit underscores the potential of slice-aligned intelligence as a scalable defense paradigm in next-generation mobile cores.

Future work will extend the framework to larger multi-slice deployments, explore control-plane spoofing and cross-slice injection scenarios, and integrate detection outputs with the Network Data Analytics Function (NWDAF) for real-time orchestration.

Ultimately, this research highlights that effective 5G security does not require excessive computational complexity: it requires architectural awareness, lightweight intelligence, and rigorous reproducibility.

## REFERENCES

[1] R. Dangi, A. Jadhav, G. Choudhary, N. Dragoni, M. K. Mishra, and P. Lalwani, "ML-based 5G network slicing security: A comprehensive survey," Future Internet, vol. 14, no. 4, p. 116, Apr. 2022, doi: 10.3390/fi14040116.

[2] A. Thantharate, R. Paropkari, V. Walunj, C. Beard and P. Kankariya, "Secure5G: A Deep Learning Framework Towards a Secure Network Slicing in 5G and Beyond," 2020 10th Annual Computing and Communication Workshop and Conference (CCWC), Las Vegas, NV, USA, 2020, pp. 0852-0857, doi: 10.1109/CCWC47524.2020.9031158.

[3] C. De Alwis, P. Porambage, K. Dev, T. R. Gadekallu and M. Liyanage, "A Survey on Network Slicing Security: Attacks, Challenges, Solutions and Research Directions," in IEEE Communications Surveys & Tutorials, vol. 26, no. 1, pp. 534-570, Firstquarter 2024, doi: 10.1109/COMST.2023.3312349.

[4] R. Singh, A. Mehbodniya, J. L. Webber, P. Dadheech, G. Pavithra, M. S. Alzaidi, and R. Akwafo, "Analysis of network slicing for management of 5G networks using machine learning techniques," Wireless Communications and Mobile Computing, Jun. 2022, doi: 10.1155/2022/9169568.

[5] M. Malkoc and H. A. Kholidy, "5G Network Slicing: Analysis of Multiple Machine Learning Classifiers," arXiv preprint arXiv:2310.01747, 2023.

[6] S. Michaelides, D. E. Chavez, and M. Henze, "Poster: Towards an automated security testing framework for industrial UEs," preprint arXiv:2505.16300, May 2025.

[7] S. Wijethilaka and M. Liyanage, "Survey on network slicing for Internet of Things realization in 5G networks," IEEE Communications Surveys & Tutorials, vol. 23, no. 2, pp. 957–994, Second Quarter 2021, doi: 10.1109/COMST.2021.3067807.

[8] W. Rafique, J. Rani Barai, A. O. Fapojuwo and D. Krishnamurthy, "A Survey on Beyond 5G Network Slicing for Smart Cities Applications," in IEEE Communications Surveys & Tutorials, vol. 27, no. 1, pp. 595-628, Feb. 2025, doi: 10.1109/COMST.2024.3410295.

[9] H. N. Fakhouri, S. Alawadi, F. M. Awaysheh, I. B. Hani, M. Alkhalaileh, and F. Hamad, "A comprehensive study on the role of machine learning in 5G security: challenges, technologies, and solutions," Electronics, vol. 12, no. 22, p. 4604, Nov. 2023, doi: 10.3390/electronics12224604.

[10] J. Dias, P. Pinto, R. Santos, and S. Malta, "5G network slicing: Security challenges, attack vectors, and mitigation approaches," Sensors, vol. 25, no. 13, p. 3940, Jul. 2025, doi: 10.3390/s25133940.

[11] J. Cunha, P. Ferreira, E. M. Castro, P. C. Oliveira, M. J. Nicolau, I. Núñez, X. R. Sousa, and C. Serôdio, "Enhancing network slicing security: Machine learning, software-defined networking, and network functions virtualization-driven strategies," Future Internet, vol. 16, no. 7, p. 226, Jul. 2024, doi: 10.3390/fi16070226.

[12] V. P. Singh, M. P. Singh, S. Hegde and M. Gupta, "Security in 5G Network Slices: Concerns and Opportunities," in IEEE Access, vol. 12, pp. 52727-52743, 2024, doi: 10.1109/ACCESS.2024.3386632.

[13] F. Salahdine, Q. Liu and T. Han, "Towards Secure and Intelligent Network Slicing for 5G Networks," in IEEE Open Journal of the Computer Society, vol. 3, pp. 23-38, 2022, doi: 10.1109/OJCS.2022.3161933.

[14] M. S. Abood, H. Wang, B. S. Virdee, D. He, M. Fathy, A. A. Yusuf, O. Jamal, T. A. Elwi, M. Alibakhshikenari, L. Kouhalvandi, and A. Ahmad, "Improved 5G network slicing for enhanced QoS against attack in SDN environment using deep learning," IET Communications, vol. 18, no. 13, pp. 759–777, Jun. 2024, doi: 10.1049/cmu2.12735.

[15] A. Thantharate, R. Paropkari, V. Walunj and C. Beard, "DeepSlice: A Deep Learning Approach towards an Efficient and Reliable Network Slicing in 5G Networks," 2019 IEEE 10th Annual Ubiquitous Computing, Electronics & Mobile Communication Conference (UEMCON), New York, NY, USA, 2019, pp. 0762-0767.

[16] K. Tiwari, A. K. Phulre, D. Vishnu, and D. Saravanan, "Enhancing secure key management techniques for optimised 5G network slicing security," ACIG, vol. 3, no. 2, pp. 170–210, 2024, doi: 10.60097/ACIG/200243.

[17] A. Haldorai, R. C. V, Q. S. Mahdi and P. Devasudha, "Application of AI/ML in Network-Slicing-Based

Infrastructure of the Next-Generation Wireless Networking Systems," 2023 Fifth International Conference on Electrical, Computer and Communication Technologies (ICECCT), Erode, India, 2023, pp. 1-10, doi: 10.1109/ICECCT56650.2023.10179649.

[18] J. Wang and J. Liu, "Secure and Reliable Slicing in 5G and Beyond Vehicular Networks," in IEEE Wireless Communications, vol. 29, no. 1, pp. 126-133, February 2022, doi: 10.1109/MWC.001.2100282.

[19] S. Gao, R. Lin, Y. Fu, H. Li, and J. Cao, "Security threats, requirements and recommendations on creating 5G network slicing system: A survey," *Electronics*, vol. 13, no. 10, p. 1860, May 2024, doi: 10.3390/electronics13101860.

[20] J. S. B. Martins et al., "Enhancing network slicing architectures with machine learning, security, sustainability and experimental networks integration," IEEE Access, vol. 11, pp. 69144–69163, 2023, doi: 10.1109/ACCESS.2023.3292788.